\documentclass[iop,numberedappendix,twocolappendix]{emulateapj}

\usepackage{float} 
\usepackage{url}
\usepackage{natbib}
\usepackage[bf]{subfigure}
\usepackage[varg]{txfonts}
\usepackage{graphicx}

\usepackage{enumerate}
\usepackage{color}
\usepackage{hyperref}
\usepackage{rotating}
\usepackage{lineno}

\def\,{\thinspace}
\def\lsim{\mathrel{\raise .4ex\hbox{\rlap{$<$}\lower 1.2ex\hbox{$\sim$}}}}
\def\gsim{\mathrel{\raise .4ex\hbox{\rlap{$>$}\lower 1.2ex\hbox{$\sim$}}}}

\def\simprop{\mathrel{\raise .4ex\hbox{\rlap{$\propto$}\lower 1.2ex\hbox{$\sim$}}}}
\def\deg{\ifmmode^\circ\else$^\circ$\fi}
\def\pdeg{\ifmmode $\setbox0=\hbox{$^{\circ}$}\rlap{\hskip.11\wd0 .}$^{\circ}
          \else \setbox0=\hbox{$^{\circ}$}\rlap{\hskip.11\wd0 .}$^{\circ}$\fi}
\def\arcs{\ifmmode {^{\scriptstyle\prime\prime}}
          \else $^{\scriptstyle\prime\prime}$\fi}
\def\arcm{\ifmmode {^{\scriptstyle\prime}}
          \else $^{\scriptstyle\prime}$\fi}
\newdimen\sa  \newdimen\sb
\def\parcs{\sa=.07em \sb=.03em
     \ifmmode \hbox{\rlap{.}}^{\scriptstyle\prime\kern -\sb\prime}\hbox{\kern -\sa}
     \else \rlap{.}$^{\scriptstyle\prime\kern -\sb\prime}$\kern -\sa\fi}
\def\parcm{\sa=.08em \sb=.03em
     \ifmmode \hbox{\rlap{.}\kern\sa}^{\scriptstyle\prime}\hbox{\kern-\sb}
     \else \rlap{.}\kern\sa$^{\scriptstyle\prime}$\kern-\sb\fi}

\def\pras{\sa=.08em \sb=.03em
     \ifmmode \hbox{\rlap{.}\kern\sa}^{\mathrm{s}}\hbox{\kern-\sb}
     \else \rlap{.}\kern\sa$^{\mathrm{s}}$\kern-\sb\fi}

\def\GHz{\ifmmode $\,GHz$\else \,GHz\fi}
\def\MJysr{\ifmmode \,$MJy\,sr\mo$\else \,MJy\,sr\mo\fi}
\def\microns{\ifmmode \,\mu$m$\else \,$\mu$m\fi}
\def\micron{\microns}
\def\kms{\ifmmode $\,km\,s$^{-1}\else \,km\,s$^{-1}$\fi}
\providecommand{\sorthelp}[1]{}

\newcommand{\beq}{\begin{equation}}
\newcommand{\eeq}{\end{equation}}
\newcommand{\bdi}{\begin{displaymath}}
\newcommand{\edi}{\end{displaymath}}

\newcommand{\herschel}{\textit{Herschel}}

\newcommand{\abreak}{$A_{\mathrm{V}}$(SF)}

\newcommand{\plogt}{P_{\ell 10}}
\newcommand{\aslope}{s}
\newcommand{\atslope}{a}
\newcommand{\fmass}{f}
\newcommand{\flogs}{c}

\newcommand{\ntot}{n_p}

\newcommand{\wo}{w}
\newcommand{\wt}{ww}

\newcommand{\hpeter}[1]{}
\newcommand{\halana}[1]{}

\slugcomment{Accepted in the Astrophysical Journal}
\shorttitle{W3 GMC: Clues to High-mass Star Formation}
\shortauthors{Rivera-Ingraham et al.}

\begin{document}


\title{\textit{Herschel} Observations of the W3 GMC (II):\\ 
Clues to the Formation of Clusters of High-Mass Stars}

\author{A.~Rivera-Ingraham\altaffilmark{1,2},
        P.~G.~Martin\altaffilmark{3},
        D.~Polychroni\altaffilmark{4}, 
        N.~Schneider\altaffilmark{5,6,7}, 
        F.~Motte\altaffilmark{5},
        S.~Bontemps\altaffilmark{6,7}, 
        M.~Hennemann\altaffilmark{5}, 
	A.~Men'shchikov\altaffilmark{5},
        Q.~Nguyen Luong\altaffilmark{3,8}, 
	A.~Zavagno\altaffilmark{9},
        Ph.~Andr\'{e}\altaffilmark{5}, 
        J.-Ph.~Bernard\altaffilmark{10,11},
        J.~Di~Francesco\altaffilmark{12,13},
        C.~Fallscheer\altaffilmark{12,13},
        T.~Hill\altaffilmark{5},
	V.~K\"{o}nyves\altaffilmark{5,14}, 
        A.~Marston\altaffilmark{2}, 
        S.~Pezzuto\altaffilmark{15}, 
        K.~L.~J.~Rygl\altaffilmark{16}, 
        L.~Spinoglio\altaffilmark{15}, 
        G.~J.~White\altaffilmark{17,18}
}


\altaffiltext{1}{Department of Astronomy and Astrophysics, University of Toronto, 50 St. George Street, Toronto, ON M5S~3H4, Canada}
\altaffiltext{2}{European Space Astronomy Centre (ESA/ESAC), P.O. Box 78, E-28691 Villanueva de la Canada, Madrid, Spain; \email{alana.rivera@esa.int}}
\altaffiltext{3}{Canadian Institute for Theoretical Astrophysics, University of Toronto, 60 St. George Street, Toronto, ON M5S~3H8, Canada} 
\altaffiltext{4}{Department of Astrophysics, Astronomy and Mechanics, Faculty of Physics, University of Athens, Panepistimiopolis, 15784 Zografos, Athens, Greece}
\altaffiltext{5}{Laboratoire AIM Paris-Saclay, CEA/IRFU - CNRS/INSU - Universit\'e Paris Diderot, Service d’Astrophysique, CEA-Saclay, F-91191, Gif-sur-Yvette Cedex, France}
\altaffiltext{6}{Universit\'e Bordeaux, LAB, UMR 5804, F-33270 Floirac, France}
\altaffiltext{7}{CNRS, LAB, UMR 5804, F-33270 Floirac, France}
\altaffiltext{8}{National Astronomical Observatory of Japan, Chile Observatory, 2-21-1 Osawa, Mitaka, Tokyo 181-8588, Japan}
\altaffiltext{9}{Aix Marseille Universit\'e, CNRS, LAM (Laboratoire d'Astrophysique de Marseille) UMR 7326, 13388, Marseille, France}
\altaffiltext{10}{Universit\'{e} de Toulouse; UPS-OMP; IRAP; Toulouse, France}
\altaffiltext{11}{CNRS; IRAP; 9 Av. colonel Roche, BP 44346, F-31028 Toulouse cedex 4, France}
\altaffiltext{12}{National Research Council Canada, Herzberg Institute of Astrophysics, 5071 West Saanich Road, Victoria, BC, V9E 2E7, Canada}
\altaffiltext{13}{Department of Physics and Astronomy, University of Victoria, PO Box 355, STN CSC, Victoria, BC, V8W 3P6, Canada}
\altaffiltext{14}{IAS, CNRS/Universit\'{e} Paris-Sud 11, F-91405 Orsay, France}
\altaffiltext{15}{INAF-Istituto di Astrofisica e Planetologia Spaziali, via Fosso del Cavaliere 100, I-00133 Rome, Italy}
\altaffiltext{16}{Istituto di Radioastronomia (INAF-IRA), Via P. Gobetti 101, 40129 Bologna, Italy}
\altaffiltext{17}{Department of Physical Sciences, The Open University, Milton Keynes MK7 6AA, UK}
\altaffiltext{18}{RALspace, The Rutherford Appleton Laboratory, Chilton, Didcot OX11 0NL, UK}



\begin{abstract}

The W3 GMC is a prime target for investigating the formation of high-mass stars 
and clusters.  
This second study of W3 within the HOBYS Key Program provides
a comparative analysis of subfields within W3 to further constrain the 
processes leading to the observed structures and stellar
population.
Probability density functions (PDFs) and cumulative mass distributions
(CMDs) were created from dust column density maps, quantified as
extinction $A_{\mathrm{V}}$.  The shape of the PDF, typically
represented with a lognormal function at low $A_{\mathrm{V}}$
``breaking" to a power-law tail at high $A_{\mathrm{V}}$, is
influenced by various processes including turbulence and
self-gravity.  
The breaks can also be identified, often more
readily, in the CMDs.
The PDF break from lognormal
(\abreak\ $\approx 6-10$\,mag) appears to shift 
to higher $A_{\mathrm{V}}$ by stellar feedback,
so that high-mass star-forming regions
tend to have higher PDF breaks.
A second break at $A_{\mathrm{V}} > 50$\,mag traces structures formed 
or influenced by a dynamic process.  Because such a
process has been suggested to drive high-mass star formation in W3, this
second break might then identify regions with potential for hosting
high-mass stars/clusters.
Stellar feedback appears to be a major mechanism driving the local
evolution and state of regions within W3.  A high initial star
formation efficiency in a dense medium could result in a
self-enhancing process, leading to more compression and favorable
star-formation conditions (e.g., colliding flows),
a richer stellar content, and massive stars. This scenario would 
be compatible with the ``convergent constructive feedback" model
introduced in our previous \herschel\ study.

\end{abstract}

\keywords{ISM: dust, extinction -- ISM: individual (Westerhout 3) -- Infrared: stars -- Stars: formation -- Stars: early-type}

\maketitle


\section{Introduction}\label{sec:intro}

\begin{figure*}[ht]
\centering
\includegraphics[scale=0.5,angle=0]{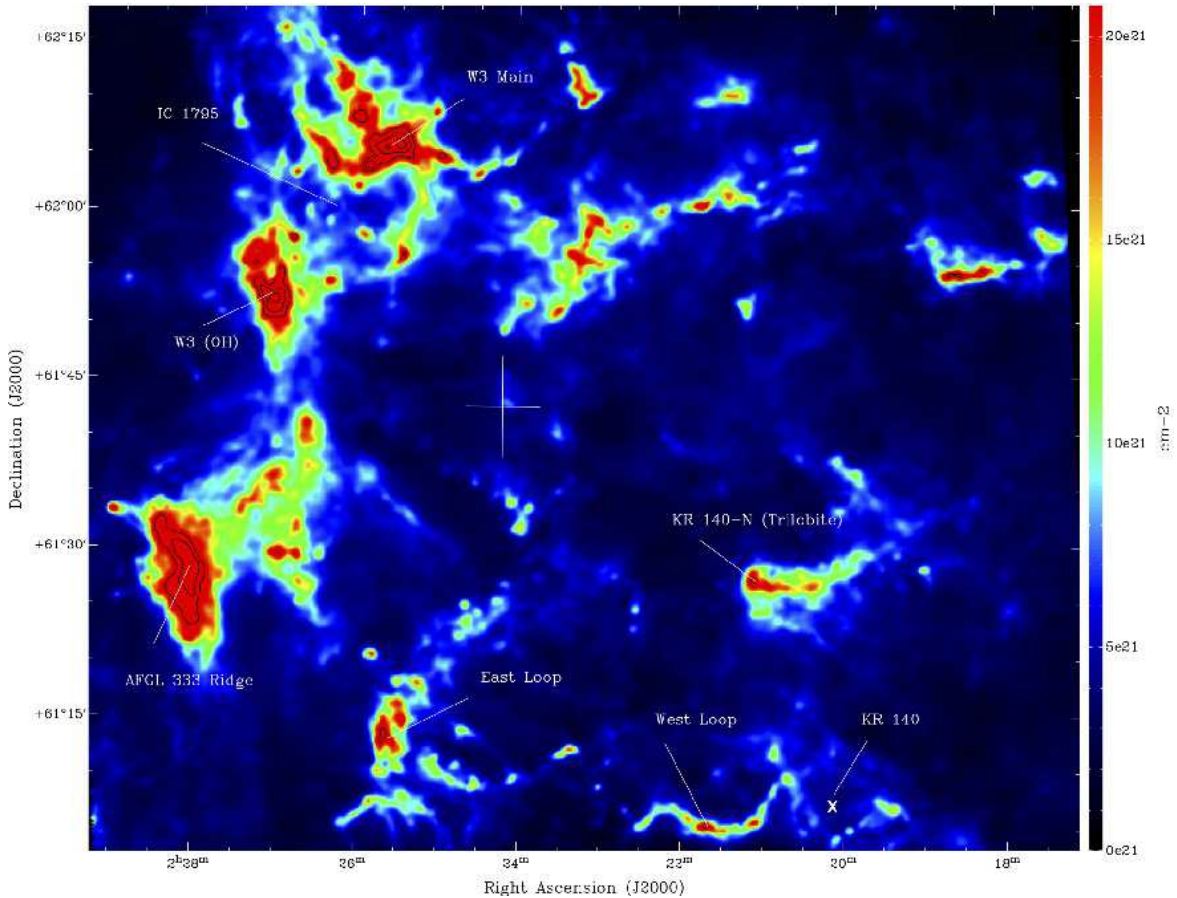}
\caption{
$N_{\mathrm{H}_2}$ column density map of the W3 GMC corrected for
contributions from foreground/background material. 
Labels mark prominent features in the cloud. The white cross
marks the intersection of the four subfields in W3; counterclockwise
from upper left: W3 Main/(OH), AFGL 333, KR 140, and W3~NW. 
The position of VES 735 is marked with a white X. The HDL is
the dense region at the east comprising W3 Main, W3 (OH), and the AFGL
333 Ridge. Contours supplement the colorbar at the highest column
densities, at $N_{\mathrm{H}_2} \approx [30, 60, 200]
\times10^{21}$\,cm$^{-2}$.
}
\label{fig:fields}
\end{figure*}

The Giant Molecular Cloud (GMC) W3 is rich in high-mass star activity
(e.g., \citealp{megeath2008}) and its relatively proximity 
($\sim2$\,kpc; e.g., \citealp{hachisuka2004}; \citealp{xu2006};
\citealp{navarete2011}) makes it a prime target for the study of
cluster and high-mass star formation.  W3 contains high-mass stars in
various evolutionary stages (e.g., \citealp{tieftrunk1997}). The
eastern high density layer (HDL) neighboring W4 contains the most
active star-forming sites: W3 North, W3 Main, W3 (OH), and AFGL
333. Activity in most of these regions might have been triggered by
nearby clusters and high-mass stars (e.g., \citealp{oey2005}). More
localized high-mass star formation is found in the western fields
(e.g., the KR 140 \ion{H}{2} region), which show indications of a more
quiescent or isolated evolution with sporadic or sequential periods of
star formation (e.g., \citealp{rivera2011}; see \citealp{rivera2012},
\citealp{rivera2013} (Paper I), and \citealp{megeath2008} for a
detailed description of the cloud and a review of recent literature).
The most prominent regions in W3 have been labeled in
Figure~\ref{fig:fields}, which shows the column density map 
at a resolution of $\sim36$\arcsec\ produced with \herschel\footnote{
\herschel\ is an ESA space observatory with science instruments
provided by European-led Principal Investigator consortia and with
important participation from NASA and the CSA.
} data (Paper I).

In Paper I we used the column density map to investigate the nature of
the most massive and highest column density structures. We found these structures 
to be most likely the result of feedback by high-mass stars, the
combined, convergent effect of which might ultimately be responsible
for the formation of the most massive Trapezium-like systems and
clusters.

In this paper we present the second part of our \textit{Herschel}
analysis of the W3 GMC (\citealp{rivera2012}; \citealp{rivera2013}).
Results below support the conclusion from Paper I that local stellar
feedback appears to be a major player not only in the formation of
clusters of high-mass stars but also in the overall process
determining the characteristics and local evolution within a GMC.

In Section~\ref{sec:data}, we provide a brief introduction to the
\textit{Herschel} datasets and analysis techniques.
Section~\ref{sec:structures} introduces the probability density
functions (PDFs) and cumulative mass distributions (CMDs)
and the procedure for selecting the structures associated with star
formation.  Appendix~\ref{pdf_uncorrected} discusses the effects of
foreground/background (non-GMC) material on the analysis and
Appendix~\ref{pdf_understand} describes some details of the
interrelationship between the PDFs and CMDs used.
Section~\ref{sec:pdf} embarks on the interpretation of structure.  
The results of our comparative analysis of the subfields in W3 are
included in Section~\ref{sec:star_formation}.
Section~\ref{sec:tracing} presents new evidence from the PDFs that
could be used to trace and constrain the birthplaces of clusters of
high-mass stars in a given region.
We conclude in Section \ref{sec:conclusion} with a summary of the key
findings related to cloud structure and star formation properties of
this cloud and how they relate to the results presented in our
previous \herschel\ study of W3 regarding the high-mass star
formation process.
%


\section{Data Processing and Maps of Column Density}\label{sec:data}

The W3 GMC was observed with \herschel\ \citep{pilbratt2010} as part
of the HOBYS\footnote{
http://www.herschel.fr/cea/hobys/en/
}
Key Programme (\herschel\ imaging survey of OB Young Stellar objects;
\citealp{motte2010}).
In Paper I we presented the \textit{Herschel} observations (SPIRE/PACS
parallel scan ObsIDs: 1342216019, 1342216020; bright mode SPIRE ObsIDs
(for saturation correction): 1342239797, 1342239796), ancillary data, and
the techniques adopted to create and analyze maps
of dust optical depth and temperature for this cloud. We recall that
these maps were made by assuming a constant dust temperature along the
line of sight and fitting spectral energy distributions (SEDs)
pixel-by-pixel using the \herschel\ dust emission maps at wavelengths
$\ge160$\,\micron\ convolved to the resolution of the $500$\,\micron\
map ($\sim$36\arcsec; $\sim$0.35\,pc at a distance of 2\,kpc).  To
describe the SED we assumed a fixed dust emissivity index of
$\beta=2$.

The optical depth and dust temperature maps were also corrected for
the effects of emission from dust in the foreground/background of
W3. This process removed the non-GMC components to reveal more closely
the properties of the interstellar medium within the GMC. The correction was
accomplished by estimating the contribution to the emission $I_{\nu}$ at each
\herschel\ band from dust traced by atomic and molecular gas (H~I and
CO emission, respectively); see \citet{rivera2012} and Appendix B of
Paper I for a detailed description of the steps and assumptions associated with 
this technique.  
This work uses these foreground/background
interstellar medium (ISM)-corrected maps as the default for our
analysis.


\subsection{Alternative Representations of Column Density}\label{ssec:column}

To convert the observable, the dust optical depth, into gas column
density, $N_{\mathrm{H}_2}$ (Figure~\ref{fig:fields}), we assumed a
dust opacity of 0.1~cm$^2$~gm$^{-1}$ at 1~THz and a mean atomic weight
per molecule of $\mu=2.33$.  The latter may be as high as 2.8
\citep{roy2013} but this slight inconsistency was retained here so
that the column densities could be directly compared with results in
other \herschel\ and HOBYS fields based on the same assumptions (e.g.,
\citealp{andre2010}; \citealp{motte2010}; \citealp{hill2011};
\citealp{hill2012}).  While $\beta$ is not definitively known in
molecular clouds and recent work points to opacity variations that
should be considered \citep{roy2013,planck2013-p06b}, none of these
systematic uncertainties in scale should impact our results
significantly, nor is such an investigation within the scope of this
paper.

Many results in the literature relevant to the topic of this paper are
given in terms of magnitudes of dust extinction $A_{\mathrm{V}}$
although that is not usually directly observable.  For the purposes
here we adopted the common approach of transforming $N_{\mathrm{H}_2}$
to $A_{\mathrm{V}}$ using $N_{\mathrm{H}_2}=0.94\times10^{21}\,
A_{\mathrm{V}}$\,cm$^{-2}$, although this is calibrated only for lower
column densities and largely atomic lines of sight \citep{bohlin1978}.
Converting directly from submillimeter dust optical depth to
$A_{\mathrm{V}}$ would seem more desirable/less contrived and indeed
the challenges of doing this have been discussed
\citep{planck2013-p06b}.  However, again our conclusions here should
not depend on these precise details.

For the corrected maps applying to the GMC material at a common
distance of 2~kpc, we can convert the column
density as parameterized by $A_{\mathrm{V}}$ into a mass per pixel: $M_p = \fmass\, A_V$.  For
the 9\arcsec\ pixels used and the assumptions described above (distance to W3 and mean atomic weight per molecule) 
we find $\fmass = 0.13$\,M$_\odot$\,mag$^{-1}$.  


\section{Methodology: Identification and Characterization of Star-Forming Structures}\label{sec:structures}

To characterize the density structures in W3 we created PDFs of the
column density maps and from these mass distributions.

\begin{figure}[ht]
\centering
\includegraphics[scale=0.49,angle=0]{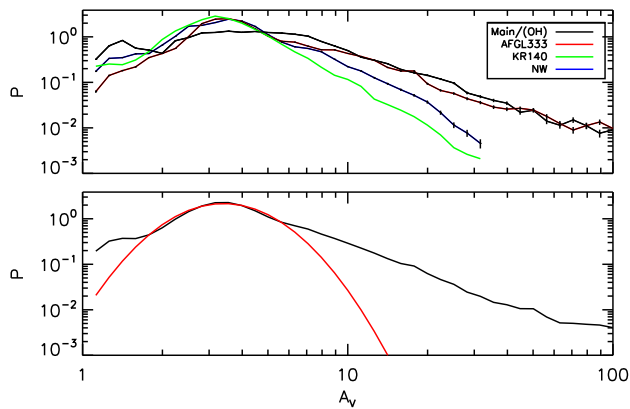}
\caption{
Probability Density Functions (PDFs) in W3.  The logarithmized version
binned in $\log A_{\mathrm{V}}$ is plotted.  Top: PDFs for each of
four fields in W3.  Errors bars are shown here as a reference.
Bottom: Global PDF for the entire W3 GMC with a fitted lognormal
function (red) and power-law tail.  
}
\label{fig:pdf_combined_ism2}
\end{figure}

PDFs quantify the fraction (or the probability $p$) of material in the
cloud having a column density in the range $N_{\mathrm{H}_2}$ and
$N_{\mathrm{H}_2}+\Delta$$N_{\mathrm{H}_2}$.  PDFs have been
used extensively as analytical tools for describing the distribution
of mass in regions of low-mass star formation (e.g.,
\citealp{froebrich2010}; \citealp{kainulainen2009}) and in regions of
high-mass star formation as well (e.g., HOBYS studies: \citealp{hill2011};
\citealp{schneider2012}; \citealp{hill2012}; \citealp{russeil2013}).

The lower panel of Figure \ref{fig:pdf_combined_ism2} shows the PDF
for the entire W3 GMC.  What is plotted is a ``logarithmized" version
of the PDF, $P \equiv \plogt$, where the independent variable (for the
binning) is now $\log A_{\mathrm{V}}$ (base 10).

The shape of PDFs is not always straightforward to interpret, as
various physical processes imprint on its structure.  For example,
interstellar turbulence likely determines the lognormal distribution
of low column-densities (see theoretical work of e.g.,
\citealp{klessen2000}, and observations of \citealp{kainulainen2009}
for all cloud types: quiescent, low-mass, and high-mass star-forming
clouds).  Following the approach used in previous studies, the main
peak in the above PDF and those below was fitted with a lognormal
distribution.  All fits were carried out using a non-linear
least-squares minimization {\sc idl} routine based on {\sc mpfit}
\citep{mpfit} and the result plotted along with the underlying data.

External compression can cause a broadening of the PDF, as predicted
in models of \citet{federrath2013} and seen in Orion B
\citep{schneider2013} and in clouds associated with \ion{H}{2} regions
\citep{tremblin2014}.  The shape of the PDF might therefore be used to
infer concrete basic physical properties of the cloud and individual
structures whose material is traced by the PDF (e.g.,
\citealp{fischera2014}).

Of all processes, gravity plays the most important role at higher
column densities and for low-mass star-forming regions causes a
clearly defined power-law like tail in the PDF
(\citealp{kainulainen2009}; \citealp{andre2011};
\citealp{schneider2013}). High-mass star-forming regions often also
show tails at high extinction that can also be accurately modelled as
power-laws (NGC6334; \citealp{russeil2013}).  Such
high-$A_{\mathrm{V}}$ tails can also have, however, more complex
shapes as well, as has been found here within W3 and in other fields
(see also \citealp{hill2011}; \citealp{schneider2012}). Indeed,
processes such as turbulence, gravity, feedback/compression, magnetic
fields, intermittency of density fluctuations, a non-isothermal gas
phase, properties of the cloud formation processes, and even
line-of-sight effects, could lead to complex substructure in the PDF
tails, such as ``breaks" and peaks, that might not necessarily be well
represented with a simple power-law function.  In the next sections,
we will discuss in more detail how stellar feedback might also imprint
on the PDF shape.  Effects on the PDFs produced by contamination from
foreground/background material along the line of sight, avoided here,
are described briefly in Appendix~\ref{pdf_uncorrected}.

\begin{figure}[ht]
\centering 
\includegraphics[scale=0.49,angle=0]{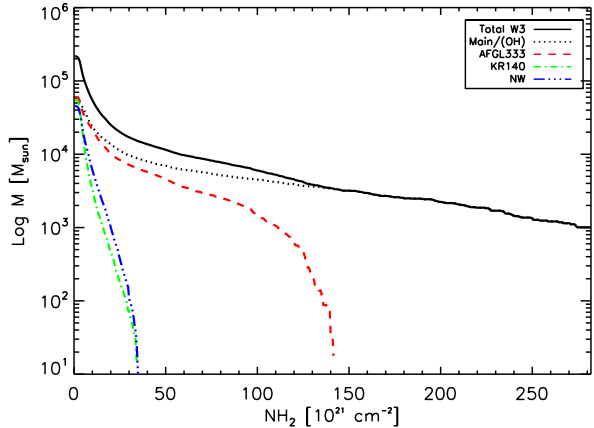}
\caption{
Cumulative mass distributions for the W3 field (black: solid line) and for each of the four fields: W3
Main/(OH) (black: dotted line), AFGL 333 (red), KR 140 (green), and W3~NW (blue).
The column densities $N_{\mathrm{H}_2}$ can be compared directly with
Figure~\ref{fig:fields}.  In terms of $A_{\mathrm{V}}$, the x-axis
range is very similar, 0 to 320\,mag.
}
\label{fig:mass_dist_ism2}
\end{figure}

Column densities theorized in other studies to be associated with
high-mass star formation are very high, for example
$\Sigma=0.7$\,g\,cm$^{-2}$ to produce a star with $M \sim
10$\,$M_{\odot}$; \citealp{km2008}, which corresponds to
$N_{\mathrm{H}_2}\sim180\times10^{21}$\,cm$^{-2}$ or $A_{\mathrm{V}}
\sim 200$ in our maps and PDFs therefrom.
Cloud column densities of this order, typically observed only towards
regions of active high-mass star formation, are rare and therefore
there are often poor statistics at the high extinction end of PDFs.
In our explorations we found that the cumulative form of the mass
distribution, the CMD (i.e., the total mass above any given magnitude;
e.g., \citealp{froebrich2010}), allows for a complementary and often
more straightforward analysis of the higher end of the PDFs.

The large differences between neighboring regions in W3 (e.g., the
dense and active HDL vs.\ the more quiescent and diffuse western
fields; \citealp{rivera2011}; Paper I) make it necessary to quantify
how in-cloud local conditions affect the star formation process.  To
this end, we have carried out an analysis of individual areas within W3
itself, these being the four fields into which we divided this
cloud in Paper I: the W3 Main/(OH), AFGL 333, KR 140, and W3~NW fields
(Figure~\ref{fig:fields}).

The PDFs of each of the four fields in W3 are given in the upper panel
of Figure \ref{fig:pdf_combined_ism2}; the characteristics of the
global PDF of W3 in the lower panel clearly depend on the
contributions from each of the four fields.

Figure \ref{fig:mass_dist_ism2} shows the corresponding CMDs, very
different for the four fields.  In this figure this is expressed in
terms of $N_{\mathrm{H}_2}$ for more direct comparison with the high
column density regions in Figure~\ref{fig:fields}.


\section{PDF Interpretation and Clues to Cloud Structure} \label{sec:pdf}

\begin{deluxetable*}{lllllll}
\tablecolumns{7} 
\tablewidth{0pc} 

\tablecaption{Breaks and Parameters\tablenotemark{a} for Linear Fits to the CMDs}

\tablehead{
\colhead{Field} & \colhead{Mass\tablenotemark{b}} & \colhead{Break1\tablenotemark{c}} & \colhead{Break2\tablenotemark{d}} & \colhead{Slope/Intercept (1)}  & \colhead{Slope/Intercept (2)} & \colhead{Slope/Intercept (3)} \\ 
& \colhead{(10$^4$\,$M_{\odot}$)} & \colhead{(mag)} & \colhead{(mag)} &&&
} 

\startdata
Main/(OH)&$6.0$&13.0&38.5&-0.0469/+4.8823&-0.0154/+4.4656&-0.0032/+3.9958\\ 
AFGL 333&$6.3$&6.0&22.5&-0.0920/+5.0419&-0.0360/+4.7095&-0.0084/+4.0903\\
KR 140&$5.9$&6.5&&-0.2120/+5.3082&-0.0945/+4.5196&\\
W3~NW&$4.8$&6.0&&-0.1465/+5.0572&-0.0844/+4.6769&
\enddata

\tablenotetext{a}{Derived from $\log M = a\, A_{\mathrm{V}}+b$, with $M$ in M$_{\odot}$.}
\tablenotetext{b}{Total (foreground/background ISM-corrected) mass of GMC in this field (see also Paper I).}
\tablenotetext{c}{\abreak\ rounded to nearest 0.5.}
\tablenotetext{d}{$A_{\mathrm{V}}$(HTB) rounded to nearest 0.5.}

\label{table:fit_params}

\end{deluxetable*}

\begin{figure}
\centering
\includegraphics[scale=0.50,angle=0]{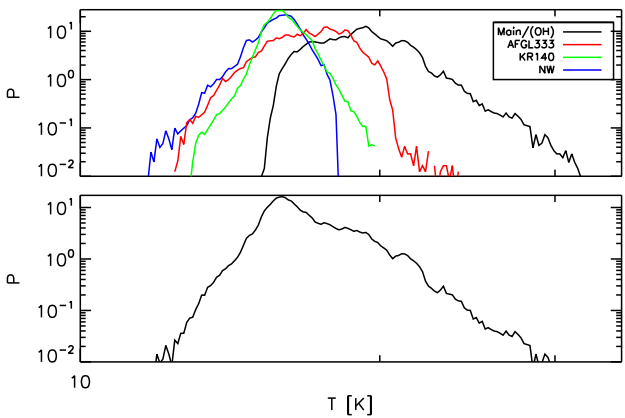}
\caption{
Like Figure~\ref{fig:pdf_combined_ism2} but for the logarithmized PDF
of dust temperature.  Top: for each field in W3. Bottom: global PDF of
the W3 GMC.
}
\label{fig:pdf_combined_ism2_temp}
\end{figure}

\begin{figure*}[ht]
\centering
\subfigure[W3 Main/(OH) field]{%
\label{fig:pdf_main_ism2}
\includegraphics[scale=0.7,angle=0,trim=1cm 0.95cm 1cm 4cm, clip=true]{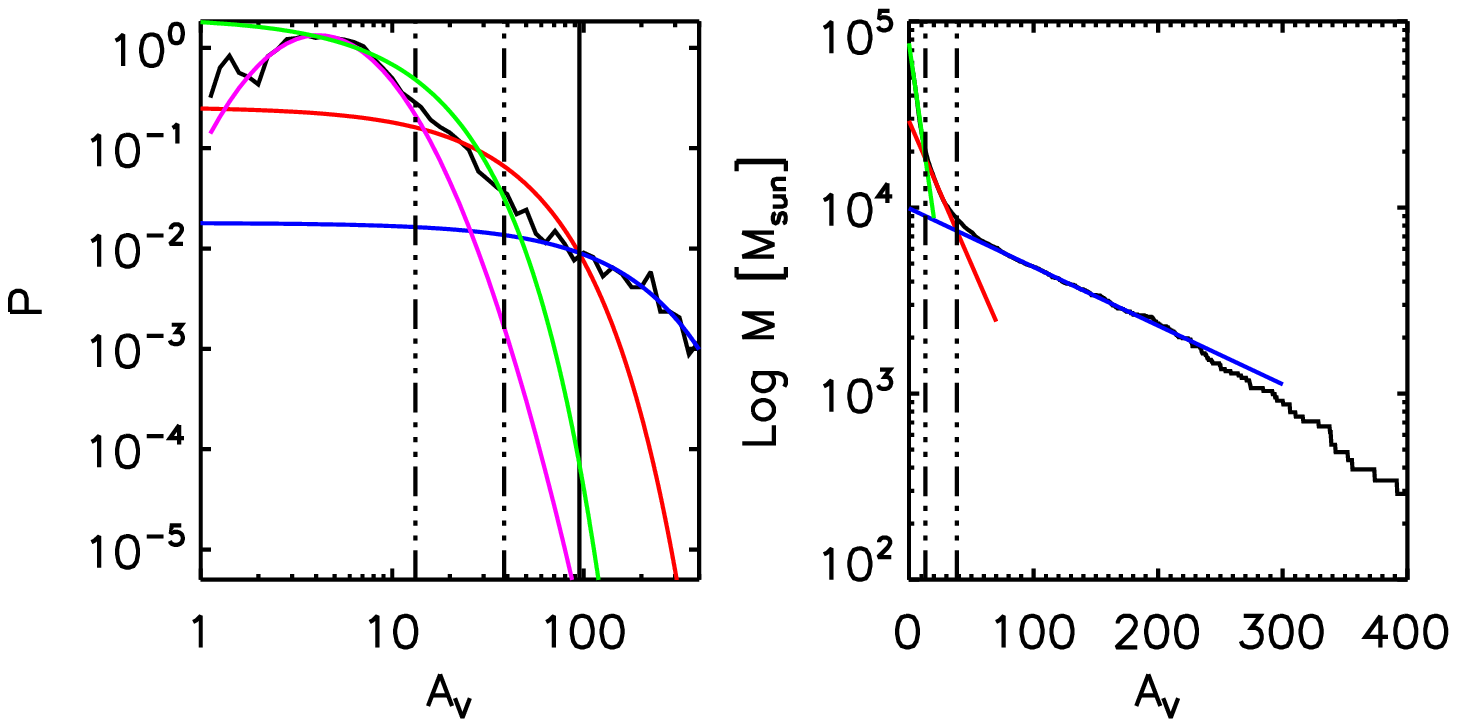}
}\\%
\subfigure[AFGL 333 field]{%
\label{fig:pdf_afgl_ism2}
\includegraphics[scale=0.7,angle=0,trim=1cm 0.95cm 1cm 4cm, clip=true]{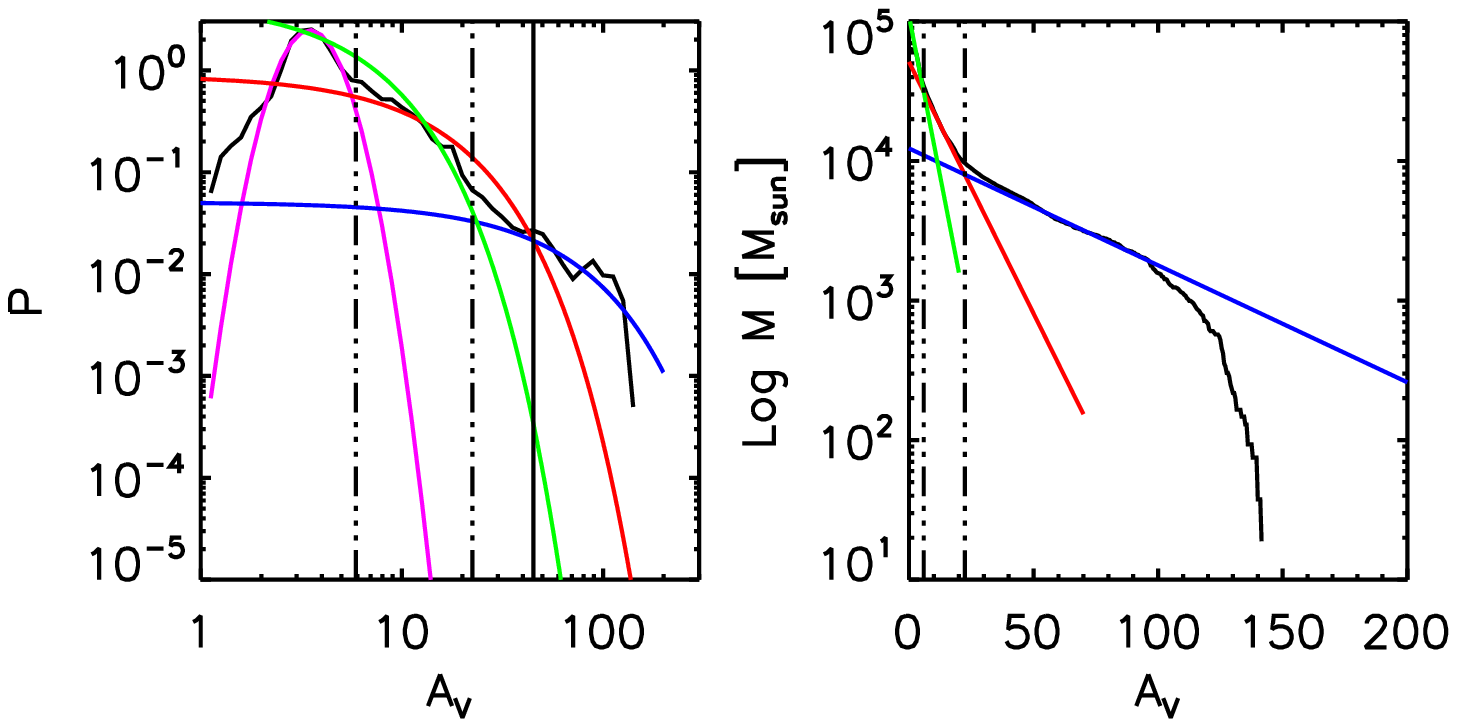}
}\\%
\caption{
PDFs (left) and CMDs (right) for the two fields in the (eastern) HDL.
CMDs: Solid green, red and blue lines are the best linear fits to the
data. Vertical dash-dotted lines mark the breaks in the distribution
(\abreak\ and $A_{\mathrm{V}}$(HTB); see text).
PDFs: Vertical dash-dotted lines and green, red, and blue curves are
those derived from the CMD (Appendix~\ref{pdf_understand}).  Magenta
curve is the best-fit lognormal function to the PDF peak. Solid
vertical lines are $A_{\mathrm{V}}$(HB), where the flatter components
start to dominate.
} 
\label{fig:pdfs1}
\end{figure*}

\begin{figure*}[ht]
\centering
\subfigure[KR 140 field]{%
\label{fig:pdf_kr_ism2}
\includegraphics[scale=0.7,angle=0,trim=1cm 0.95cm 1cm 4cm, clip=true]{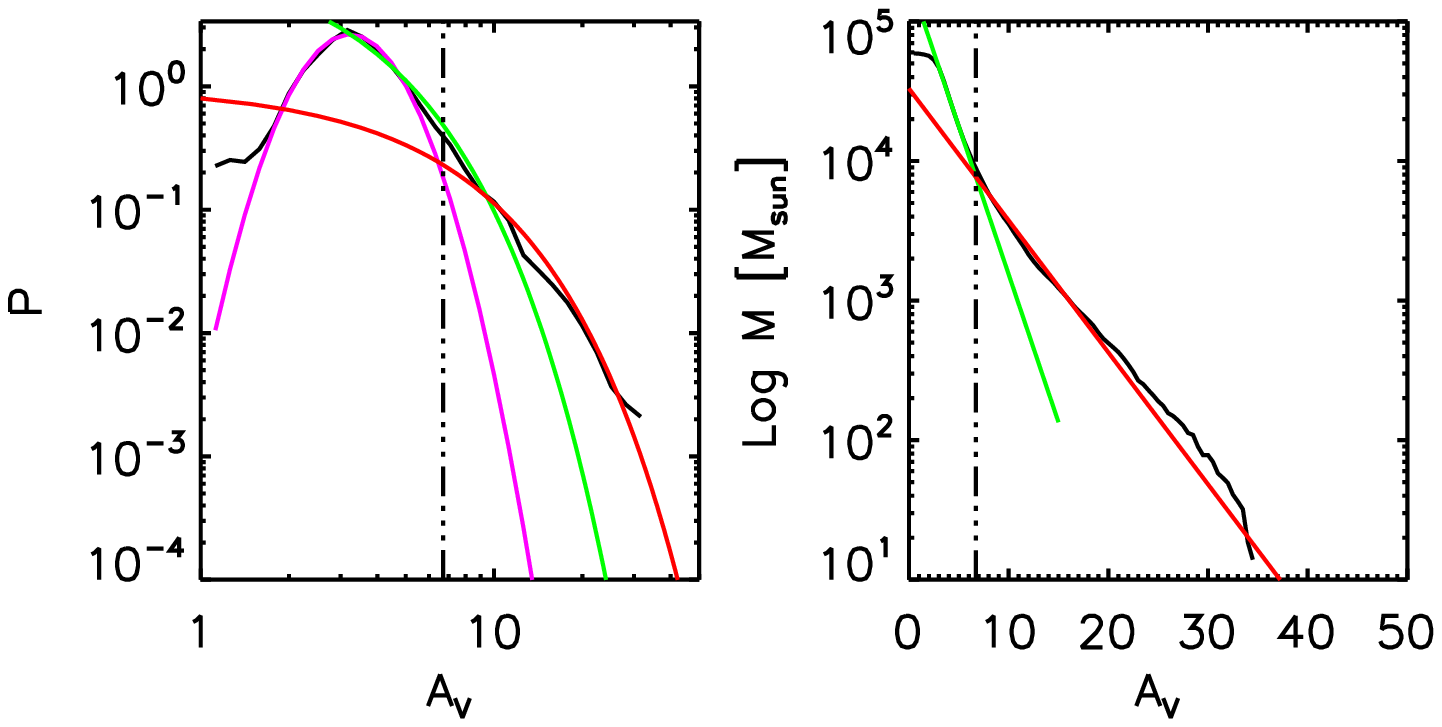}
}\\%
\subfigure[W3~NW field]{%
\label{fig:pdf_west_ism2}
\includegraphics[scale=0.7,angle=0,trim=1cm 0.95cm 1cm 4cm, clip=true]{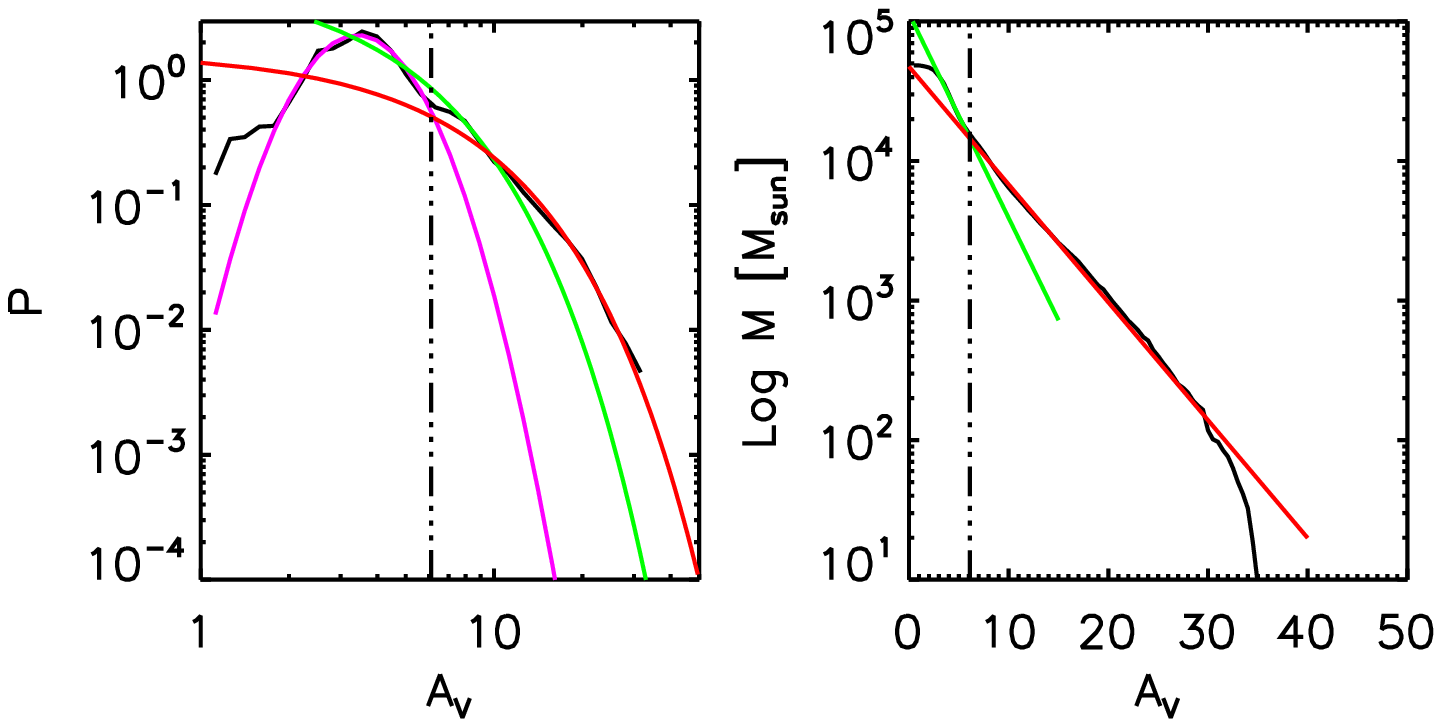}
}\\%
\caption{
Same as Figure \ref{fig:pdfs1}, but for the two western fields. These
two fields lack the flattening of the PDF at high extinction that
characterizes the HDL fields.
} 
\label{fig:pdfs2}
\end{figure*}

\begin{figure*}[ht]
\centering
\includegraphics[scale=0.7, angle=0,trim=1cm 0.95cm 1cm 4cm]{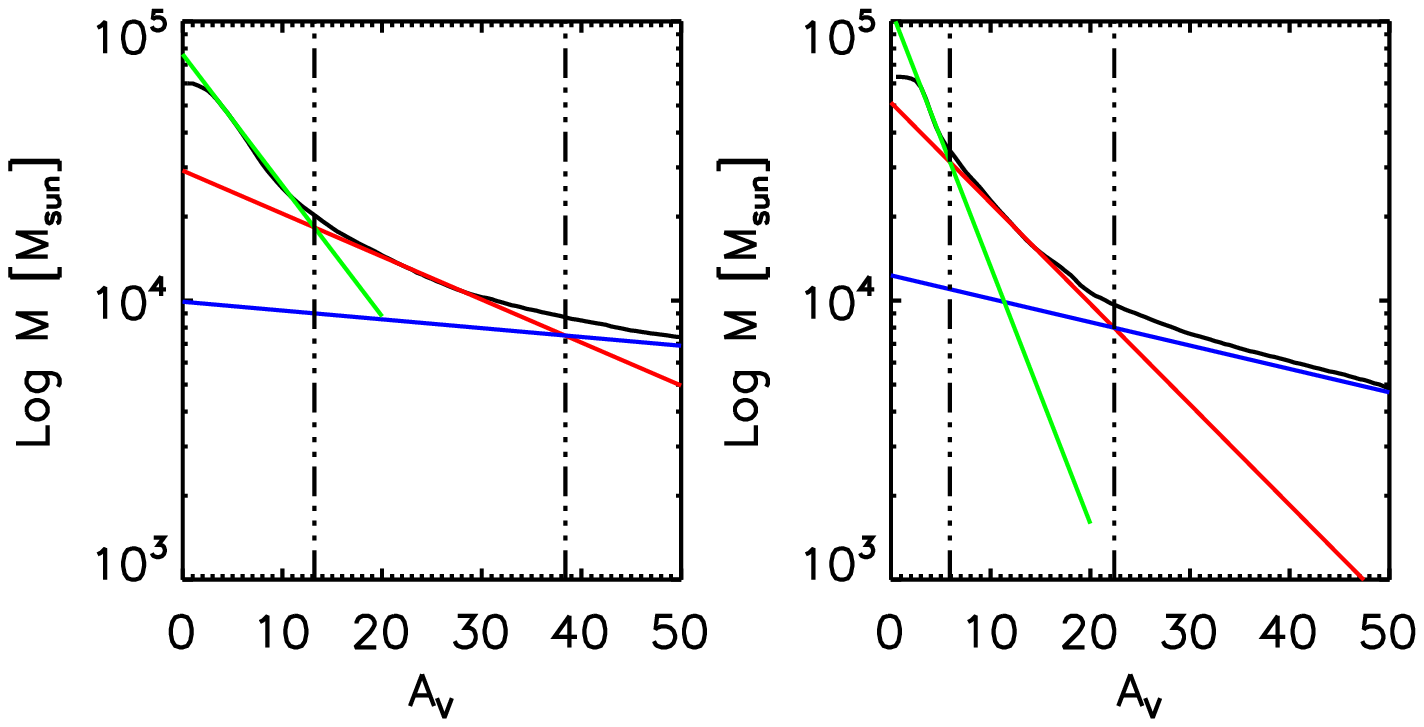}
\caption{
Expanded view of the low extinction range of the CMDs in Figure \ref{fig:pdfs1}, 
highlighting the breaks and the best linear fits to the data for W3 Main/(OH) (left), and AFGL 333 (right).
}
\label{fig:zoom}
\end{figure*}


\subsection{Analysis and Characterization of PDF Substructure} \label{ssec:break}

While various factors can influence the shape of a PDF, the column
density above which it deviates from a lognormal distribution into a
power-law (i.e., the break) has been commonly interpreted as the point
separating the turbulent medium from the star formation regime. The
PDF break can therefore be understood as the column density above
which star-forming structures start to dominate over the local
environment (e.g., \citealp{kainulainen2009}; \citealp{andre2011};
\citealp{ballesteros2011} and references therein).

This break is therefore an essential parameter for constraining the
star formation properties of individual fields.  We have found that
such breaks can be identified and characterized by fitting linear
functions to the $\log$ of the CMD. See \citealp{rivera2012} and
Appendix~\ref{pdf_understand} for a more detailed description of the
concepts and methodology used.  Although the accuracy of the estimate
of the break can be affected by the binning ($\Delta
A_{\mathrm{V}}=0.5$\,mag) and fitting procedure (uncertainties in
linear slopes are $\sim5$\,\%), a major source of systematic
uncertainty for the lowest breaks will arise from the corrections
applied to the maps to remove line-of-sight, non-GMC material, as
explained in Appendix~\ref{pdf_uncorrected}.

In our application of this approach, the division into subfields from
Paper I was used, as it separates regions with dramatically different
characteristics. Our choice of subfields was determined by unique
differences in column density and temperature distributions, stellar
activity, and stellar content. Constraining the origin of these
differences is crucial for understanding the onset of high-mass star
activity in the very specific regions of the HDL. The temperature
differences can clearly be observed, for instance, in the temperature
PDFs for each field in Figure \ref{fig:pdf_combined_ism2_temp}. The W3
Main/(OH) field is the most active and warmest, as well as the only
one with ongoing high-mass star formation. At the other extreme, the
KR 140 and W3~NW fields are cold and quiescent. The AFGL 333 has
intermediate properties, with some high-mass stars, and yet is much more
quiescent than W3 Main/(OH).

The derived values and parameters for the linear fits (slope and
intercept of the lines used to represent the mass distributions) are
given in Table~\ref{table:fit_params} and the fitted lines are shown
in the right panels of Figures~\ref{fig:pdfs1} and \ref{fig:pdfs2}. 
An expanded view of the CMDs in Figure~\ref{fig:pdfs1} has been 
included in Figure~\ref{fig:zoom}.

The Figures show that the first breaks (\abreak) derived
from the mass distributions approach those in the PDFs, while the
shapes of the components can effectively account for visual changes
(bumps) in the tails of the PDFs. The accuracy for determining the
break clearly depends on the sharpness of the transition between the
different (linear) regimes in the mass distributions (i.e., those
fitted with linear functions). A slow transition between two regimes
(i.e., not sharp, but progressive and occurring over a broader
extinction range, as observed for those in the HDL; Figure~\ref{fig:zoom}) 
would result in a higher degree of uncertainty, as the identification of the end and
starting points of the linear regimes becomes less clear. A higher
degree of uncertainty for the first break of the KR 140 field is
expected due to this effect and the fact that this field is also the
one with the most severe line-of-sight (molecular) contamination.

Prominent differences can be seen between the PDFs of the two regions
in the HDL and those of the western fields.  W3~NW and KR 140 (see Figure~\ref{fig:pdfs2}) show
simpler, well-fitted PDFs.  While a classical power-law function would
not be able to reproduce the PDF tails of the western fields, our fits
derived from the linear functions to the mass distributions can
reproduce them well.  The W3 Main/(OH) and AFGL 333 fields (see Figure~\ref{fig:pdfs1}) have much
more complex tails extending to high extinction. While our fits do not
manage to reproduce this complexity as accurately as those for the
western fields, they successfully locate the \abreak\ break and
manage to trace the overall variation of the tails with increasing
extinction. Note that, like for the western fields, the tails of these
PDFs would not be well fitted with a classic power-law function
either.

The fits to the mass distributions reveal the presence of a second
major break in the PDFs of the HDL fields (defined in this work as
$A_{\mathrm{V}}$(HTB), for ``High-extinction Transition Break"). This
break marks the transition point between the classical power-law tail
of the PDF and a ``flatter" regime, fitted in the mass distributions
with a third linear function (blue slopes; Figure~\ref{fig:pdfs1}). As
with the first break in the KR 140 field, this second transition was
found to be considerably wide and progressive (i.e., characterized by
a smooth, slow transition) for both HDL fields, therefore resulting in
a higher degree of uncertainty when determining the true transition
point. The regions traced by these intermediate (transition)
extinctions are identified with the structures and shells in W3
Main/(OH) and the AFGL 333 fields, which host the most massive ridges
and clumps in the W3 GMC.

The actual break at which the flat regime at high extinctions starts
to dominate in the PDF ($A_{\mathrm{V}}$(HB)) is marked by the point
where the mass distribution is properly described by the third linear
function, whose parameters have been included in Table
\ref{table:fit_params}. This extinction is also coincident with the point where
the two linear fits (blue and red) in the mass distributions intersect
in the PDF (Figure~\ref{fig:pdfs1}). These points are
$A_{\mathrm{V}}$(HB)$\sim45$\,mag and $95$\,mag for AFGL 333 and W3
Main/(OH), respectively. The presence and nature of this possible
break is important due to its link with high column density material,
including high-mass star-forming regions.


\subsection{Interpretation and Clues to Cloud Structure}
 
While all fields have comparable total mass, those in the HDL have
distributions of material at higher extinctions that distinguish them,
not only from the KR 140 and W3~NW fields, but also between themselves
(e.g., based on Kolmogorov-Smirnov (KS) probability
tests). Considering also their dramatic differences in (high-mass)
stellar activity, characterization of the different $A_{\mathrm{V}}$
regions, with the aid of the PDFs, is essential for understanding the
processes driving current star formation and the history that led to
these (successful) conditions.  Below, we discuss each
$A_{\mathrm{V}}$ range in turn.


\subsubsection{Low $A_{\mathrm{V}}$ Range: Constraining Environmental Conditions}

Compared with other high-mass star-forming regions (e.g., Rosette
cloud; \citealp{schneider2012}), the properties of the PDFs/mass
distributions of W3 indicate that material with extinction
$A_{\mathrm{V}}\sim3$\,mag comprises a typical GMC environment, or
\textit{common plateau}, on which the structures with star-forming
potential are observed. Note that the critical extinction for core
formation of $A_{\mathrm{V}}\sim8$\,mag \citep{andre2010} relies on a filamentary environment, so the
above estimate represents the environment of the filament itself.

Excluding W3 Main/(OH), the only field with confirmed on-going
high-mass star formation (e.g., HC \ion{H}{2} regions), the actual
transition point to the gravity-dominated regime in W3 is found to be
$<$\abreak$>\approx6$\,mag.  More than $90\,\%$ of the Class
0/I YSO population \citep{rivera2011} is contained above the 
\abreak\ of a given field (except in KR 140, where
$\sim30\,\%$ of the population is below the \abreak). The
PDF breaks are therefore suitable limits for identifying the major
current and potential star-forming sites in the GMC, separating these
from typical environmental column densities in the cloud. The amount
of star-forming cloud mass in each field, based on these limits, is
shown in Table \ref{table:sf_parameters_ism}.

Similarly, we can also define typical environmental temperatures, or
$T_{\mathrm{env}}$, for each field. The temperatures associated with
material below the \abreak\ coincide with the peaks of the
temperature PDFs shown in Figure \ref{fig:pdf_combined_ism2_temp};
i.e., $T\approx19.5$\,K, $17.5$\,K, $16.0$\,K, $16.0$\,K for W3
Main/(OH), AFGL 333, KR 140, and W3~NW, respectively. Note, however,
that a key property of the youngest high-mass star-forming regions in
W3 is their association with high-column density material with
temperatures highly above environmental (Paper I). Exclusion of such
warm material when identifying young regions would clearly
underestimate the true star-forming potential of the GMC, especially
in terms of high-mass stars.

\begin{deluxetable}{lllll}
\tablecolumns{5} 
\tablewidth{0pc} 

\tablecaption{Characteristics of On-going and Potential Star-forming Structures}

\tablehead{
\colhead{\phm{\wt}$M$\phm{\wt}\phm{\wo}} & \colhead{\%\tablenotemark{a}\phm{\wo}} & \colhead{$<T>$\phm{\wo}} & \colhead{Area\phm{\wo}} & \colhead{\%\phm{\wo}}\\
\colhead{($10^3$ M$_{\odot}$)} && \colhead{(K)} & \colhead{(pc$^2$)} &
}

\startdata
\sidehead{W3 Main/(OH)}
\tablenotemark{b}$20.5\pm0.12$&34.2&$21.1$&39.8&7.6\\
\tablenotemark{c}$4.9\pm0.04$&8.2&$18.3$&11.1&2.1\\
\sidehead{AFGL333}
$34.1\pm0.09$&53.9&$17.1$&151.1&24.7\\
$19.8\pm0.07$&31.3&$15.5$&80.4 &13.1\\
\sidehead{KR140}
$9.5\pm0.03$ &16.2&$15.5$&59.4 &6.6\\
$7.3\pm0.03$ &12.4&$15.1$&42.9 &4.7\\
\sidehead{W3~NW}
$15.9\pm0.04$&32.8&$15.0$&97.6 &15.2\\
$14.9\pm0.04$&30.7&$14.8$&89.4 &13.9
\enddata

\tablenotetext{a}{\% of total field.}
\tablenotetext{b}{$A_{\mathrm{V}} \ge$ \abreak\ (Table \ref{table:fit_params}).}
\tablenotetext{c}{$A_{\mathrm{V}} \ge$ \abreak\ and $T\le T_{\mathrm{env}}$.}

\label{table:sf_parameters_ism}

\end{deluxetable}


\subsubsection{High $A_{\mathrm{V}}$ Range: Searching for Sites of High-Mass Star Formation}

Characterizing the nature of the second break at high $A_{\mathrm{V}}$
in the PDFs is crucial due to its possible association with regions of
high-mass star formation.  While similar breaks were also observed by
\citet{schneider2012} for the Rosette field containing the highest
column densities (reaching $A_{\mathrm{V}}\sim70$\,mag in their maps),
the AFGL 333 and W3 Main/(OH) fields reach extinctions 2-6 times this
limit, and second breaks about 2.5-4.5 that observed in Rosette. In
the following sections, we suggest that these second breaks originate
from, or are influenced by, the effects of an external dynamic
process. In the case of W3, this process is dominated by stellar
feedback and associated triggering (compression) from high-mass stars,
acting on already relatively dense star-forming structures (e.g.,
shells). Therefore, the structure of the PDFs of W3 provide additional
evidence for the convergence constructive feedback scenario of
high-mass star formation (Paper I).


\section{Discussion: 
Favorable Conditions for High-Mass Star Formation}\label{sec:star_formation}

The W3 GMC offers a unique opportunity to investigate star formation
under different environmental conditions, from high density structures
with high stellar feedback (eastern HDL), to more diffuse and
quiescent (western) regions with localized star formation
\citep{rivera2011}.  In the following sections we compare the fields
in W3 to constrain the conditions that have led to the (rare) onset of
high-mass star formation in this GMC.


\subsection{The Environmental Factor \abreak\ and the Role of Stellar Feedback in Determining Local (in-Cloud) and Global Evolution}

The HDL, hosting the only young high-mass population in W3, comprises
up to $\sim70\,\%$ of the dense (above environmental limit; 
\abreak) material in the GMC.  For comparison, W3 Main/(OH)
and AFGL 333 contain up to $\sim2.5$ times more mass for potential
star formation than the neighboring western fields KR 140 and W3~NW.

Assuming that a third of the total mass of W3 with
$A_{\mathrm{V}}>$\abreak\ is transformed into stars
\citep{alves2007}, the HDL would therefore have a total mass fraction
involved in star formation (Mass$_{\mathrm{SF}}$; or ``MSF" in
\citealp{froebrich2010}) of $\sim15$\,\% at a resolution of
$\sim0.35$\,pc (c.f. the western fields:
$Mass_{\mathrm{SF}}\sim7.5$\,\% of their total mass
$M_{tot}\sim1.1\times10^5$\,$M_{\odot}$).  The W3 GMC as a whole shows
a Mass$_{\mathrm{SF}}\approx11.5$\,\%.

While the mass of the W3 GMC is comparable to that of Auriga 1 or
Cepheus, these clouds only have Mass$_{\mathrm{SF}}$ of just 0.19\,\%
and 0.26\,\%, respectively \citep{froebrich2010}. The maximum
Mass$_{\mathrm{SF}}$ found by \citet{froebrich2010} (in their cloud
sample) is $\sim10$\,\% (Corona Australis). Compared to these low-mass
star-forming clouds, W3 appears to have an anomalously high proportion
of mass involved in star formation, with a very high potential to form
new stars in the next $10^6$\,yrs \citep{froebrich2010} despite its
already significant ongoing star activity.

W4 shows signatures of prominent stellar activity 
that has been ongoing for at least $\sim6-20$\,Myr \citep{oey2005}. 
Regardless of the distribution of material in the W3 region prior 
to the formation of W4, the scenario of successive episodes of (high-mass) star formation 
and bubble development described by these authors suggests that W4 not only influenced 
the star formation process in W3 at the distance of the HDL, but also that 
the HDL, the densest structure in W3, originated due to this same activity 
of bubble/shell expansion, redistribution, and compression of material.
The idea of a triggered origin for the HDL (e.g., by compression) has already been 
suggested in previous studies (e.g., \citealp{moore2007}), and is 
strongly supported by extensive observational evidence. 
Its location, parallel to the W4 \ion{H}{2} region, the morphological and physical 
characteristics of the HDL presented in this work, 
as well as the W4-W3 stellar age and cluster 
distributions (e.g., \citealp{carpenter2000}; \citealp{rivera2011}), 
indicate a clear influence of W4 on the material as well as the stellar activity in 
the HDL over an extended period of time.
Based on this result, and with the HDL dominating the contribution of
dense material in the GMC, stellar feedback can therefore explain the
high Mass$_{\mathrm{SF}}$ in the HDL and therefore in W3 as a whole. Feedback
as the driver distinguishing the structures in the HDL from those in
the western fields is also supported by molecular observations
\citep{polychroni2012}.  The key issue remains as to whether or not feedback
can also explain the local differences between fields, and why
high-mass star formation is exclusive to just some particular regions
in the HDL (W3 Main/(OH)).

Theoretical models and simulations predict that stellar feedback can indeed 
have a significant impact on the star formation process. 
Depending on environmental conditions and the 
location and number of high-mass stars, feedback strongly 
disturbs cloud morphology, shifting and redistributing material and 
creating new dense regions potentially suitable for low and high-mass 
star formation alike (e.g., \citealp{whitworth1994}; \citealp{dale2007}; \citealp{walch2013}).
It can also affect the average stellar mass, by dispersing or concentrating the 
local material needed for accretion, while increasing the total number of young stars 
of a given region (e.g., \citealp{federrath2014}; \citealp{dale2015}). 
Its effects on local star formation from winds and (especially) ionization 
(e.g., \citealp{dale2013}) are, however, dependent on distance and environmental density, 
with its destructive effects being highly minimized in the densest environments 
(e.g., \citealp{dale2011}; \citealp{ngoumou2015}).
Ultimately, feedback might regulate the global star formation process at a range of 
spatial scales through the input of turbulence (e.g., \citealp{maclow2004}; \citealp{boneberg2015}) 
and a complicated balance between destructive (e.g., material dispersal) and 
constructive effects (e.g., compression, material accumulation and creation of 
dense structures; e.g., \citealp{dale2007}; \citealp{krumholz2014} and references therein). 

Based on our observations and the general theoretical predictions in the above studies, 
we suggest that stellar feedback from high-mass stars is currently 
the main factor distinguishing the observed differences in star formation between 
fields in the W3 GMC, disrupting the local environment and altering the
characteristics, onset, and evolution of the star formation process in
a given region.  This effect can be inferred from Table
\ref{table:ysos_params}, which summarises the environmental
differences between W3 YSO populations. Here we used the
$N_{\mathrm{H}_2}$ and $T$ of the pixel in the \herschel\ maps
coincident with the YSO coordinates as representative of the local
conditions in which a given YSO resides (pixel size$\sim$0.1\,pc).

\begin{deluxetable}{lll}
\tablecolumns{3} 
\tablewidth{0pc} 

\tablecaption{Average Environmental Properties of YSOs\tablenotemark{a} in the W3 GMC.}

\tablehead{
\colhead{\phm{\wt}YSOs\phm{\wt}\phm{\wo}} & \colhead{$<T_d>\phm{\wt}$} & \colhead{$<A_{\mathrm{V}}>$\phm{\wt}}\\
& \colhead{(K)} & \colhead{(mag)}
}

\startdata
\sidehead{W3 Main/(OH) + AFGL 333}
All types&18.1&$25.6$\\
Class 0/I&16.9&$56.6$\\
Class II&18.5&$14.8$\\
\sidehead{KR 140 + W3~NW}
All types&15.3&$9.1$\\
Class 0/I&14.8&$12.9$\\
Class II&15.5&$7.9$
\enddata

\tablenotetext{a}{Catalog 1, all flags from \citet{rivera2011}.}

\label{table:ysos_params}

\end{deluxetable}

The mean extinction of Class 0/I and Class II candidates in the HDL
\citep{rivera2011} is $A_{0/I-HDL}>4.4\times A_{0/I-West}$ and
$A_{II-HDL}\approx1.9\times A_{II-West}$, respectively (Table
\ref{table:ysos_params}). 
The triggering process that originated the HDL, in combination with the activity 
from the local high-mass stars in W3, have therefore provided the denser 
(and warmer) conditions suitable for the onset of the most vigorous 
and richest stellar activity currently observed in the W3 GMC.

Similarly, comparison of the local differences between Class 0/I and
Class II YSOs in a particular region ($\Delta$$<A_{\mathrm{V}}>$, or
the change in local column density with time), can constrain possible
environmental effects on YSO evolution.  
Assuming a total lifetime for
the Class 0/I (+flat SED) and Class 0/I + Class II phases of
$\sim0.9$\,Myr and $\sim2.9$\,Myr, respectively \citep{evans2009},
then for co-eval evolution (same age) Class II YSOs in the HDL could
leave or have their environment disrupted (e.g., higher external
activity) up to $\sim8$ times faster than in the western fields, whose
Class 0/I and Class II candidates co-exist in similar (cool)
environments and comparable column densities.  In a more conservative
scenario in which Class II sources in the HDL are the oldest (2.9\,Myr
old) Class II population in W3 (while those in the western fields have
just been formed; i.e., 0.9 Myr old), then Class II sources in the HDL
still dissociate from their primordial material $\sim2.5$ times faster
than those in the western fields. 
According to theoretical models, this apparently negative effect 
could result in the new stellar population in the proximity of the 
high-mass stars in the HDL being more numerous, albeit with overall lower masses 
than the more localized stellar population formed in the more quiescent 
western fields. The HDL has, however, a significant high-mass star population and massive clusters 
whose origin might also be linked to the (in this case, constructive) effects 
of triggering and external events (Paper I).

Below we summarize the observational evidence from this work supporting a feedback-driven model 
for the local evolution of regions within the W3 GMC. These results aim 
to introduce the basis of a scenario (Section~\ref{ssec:history}) in which evolution is 
intimately linked to the balance of constructive/destructive effects of the feedback mechanism, 
themselves highly dependent on local environmental density. 


\subsubsection{The HDL: Star Formation in Very High-Density Environments}

W3 Main/(OH) is the only field with ongoing (clustered) high-mass star
formation, despite the fact that the AFGL 333 is also influenced by
the activity in W4.  Considering the effects of stellar feedback on
star/structure formation, this intense star formation, as well as the other unique properties
of the W3 Main/(OH) field, might ultimately be linked to its particularly high 
degree of stellar feedback from current stars within W3 on a quite 
local parsec scale (a few arc minutes on Figure~\ref{fig:fields}),
including but certainly not exclusively high-mass stars in IC 1795 and 
those powering W3 Main.
This appears to the dominant mechanism distinguishing the current state
and evolution of the different fields. 
The key role of local high-mass stars acting already within the W3 complex, 
rather than external activity from W4, for determining the current conditions of the star-forming material 
within dense environments is based on the following observations:

1) W3 Main/(OH) has more extreme environmental conditions, with a 
$T_{\mathrm{env}}$ higher than that observed for AFGL 333. 
This would be inconsistent with the activity in W4 being a main factor 
determining the in-cloud state of the HDL, as the latter is closer (in projection), 
and more heavily irradiated by, the high-mass stellar activity in W4. 
We find that $>95\,\%$ of the AFGL 333 field is below 
the $T_{\mathrm{env}}$ of W3 Main/(OH), excluding the pillar east of the
AFGL 333 Ridge (YSO Group 7; \citealp{rivera2011}), and IRAS
02245+6115, an \ion{H}{2} region associated with a B-type star
(\citealp{hughes1982}; \citealp{straizys2010}). 

2) W3 Main/(OH) has a much more significant (and on-going) stellar
activity \citep{rivera2011}, as well as a higher disruption of the YSO
environment. About $\sim80\,\%$ of Class II sources with
$A_{\mathrm{V}}\le$ \abreak\ also have $T > T_{\mathrm{env}}$
in this field.

3) While both HDL fields reach column densities of the same order, W3
Main/(OH) also has a PDF break \abreak\ twice that of the
AFGL 333 field. This higher break selects the shell-like structures
around IC 1795, shaped by the central cluster and its (parsec-scale)
feedback from local high-mass stars \citep{rivera2011}. In addition to
gravitational effects, this PDF break could therefore be directly
influenced by the magnitude of external/active processes.

If dynamic (e.g., feedback) processes are key for anomalously
enhancing the amount of star-forming (dense) material in W3, shifting the
position of the PDF break, and driving the high-mass star/cluster
formation itself (as suggested in Paper I), then this common link
could, in addition, explain why: i) W3, a high-mass star-forming
cloud, has a higher Mass$_{\mathrm{SF}}$ than any of the low-mass
star-forming clouds in the sample of \citet{froebrich2010}, and ii)
fields classified as high-mass star-forming regions have generally a
higher \abreak\ than low-mass star-forming regions (as
observed in, e.g., \citealp{schneider2012}).


\subsubsection{The Western Fields: Star Formation in More Diffuse Environments}

\begin{figure}[ht]
\centering
\includegraphics[scale=0.8,trim=2cm 0cm 0cm 0cm]{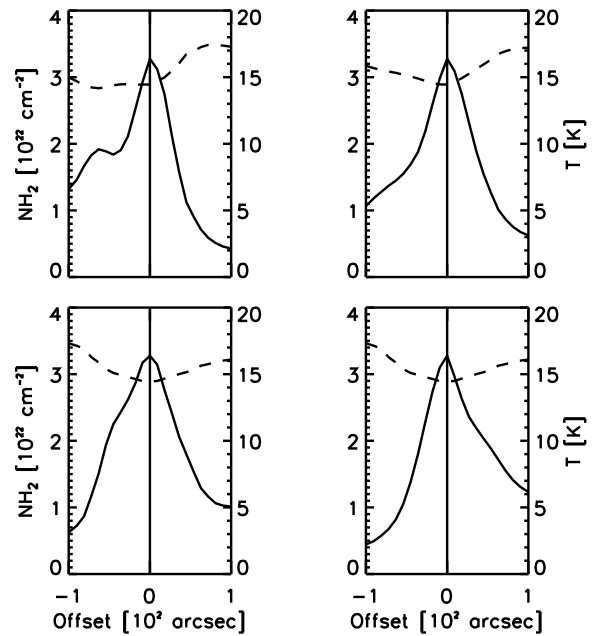}
\caption{
Profiles 200\arcsec\ ($\sim 2$\,pc) long through the column density
peak of the Trilobite centered on position RA/Dec:
$2^{\mathrm{h}}$\,21$^{\mathrm{m}}$\,5\pras3,
+61$^{\circ}$\,27\arcmin\,29\parcs1.
Both temperature (dashed line; right scale) and column density (solid
line; left scale) are shown for different orientations of the profile:
in the West-East direction (top-left); NW-SE (top-right); North-South
(bottom-left); NE-SW (bottom-right).  Black vertical line marks the
coordinate center (solid line).
}
\label{fig:profile_kr140n}
\end{figure}

In the KR 140 field, material with $A_{\mathrm{V}}>$\abreak\ 
is exclusively related to filamentary and shell-like structures; i.e.,
the Trilobite, the shell of the KR 140 \ion{H}{2}, and the West Loop.
(Figure~\ref{fig:fields}). Each of these structures has a morphology consistent
with external influence: Radiative-Driven Implosion (RDI;
\citealp{rivera2011}), the shell around the O-type star VES 735, and the
border of a cavity-like structure observed in the CO map from the
Canadian Galactic Plane Survey (CGPS; \citealp{taylor2003}) at
$v \approx49.3$\,km\,s$^{-1}$, respectively. The triggered-like origin
of these structures is supported by the (asymmetric) distribution of
their YSO population \citep{rivera2011}, and column density and
temperature profiles (e.g., asymmetric gradients). A profile example
for the Trilobite can be seen in Figure \ref{fig:profile_kr140n}.

The comparison between the two western fields resembles that between
AFGL 333 and W3 Main/(OH), albeit in a much smaller scale.  A possible
cluster in the central-southern parts of the western fields
(M. Rahman, priv.\ comm.) might have led to a period of enhanced
feedback in the KR 140 field, resulting in higher feedback and a
richer stellar content than in W3~NW; e.g., embedded clusters \citep{carpenter2000}, 
various (late) B stars \citep{voros1985}, 
and the only (confirmed) O-star (VES 735) outside
the HDL (according to
SIMBAD\footnote{http://simbad.u-strasbg.fr/simbad/}).  This affected
the temperature distribution accordingly, with KR 140 having a much
smaller proportion of ``cold" material in any given extinction range
than the W3~NW field.  

The western fields therefore provide observational evidence of a radically 
different evolutionary path with respect to those in the eastern layer. 
Based on their properties relative to those of the HDL, 
such evolutionary difference could be easily linked to differences in the level 
and type of feedback in the history of W3~NW and KR 140.
Here, structures have not 
benefitted from the primordial dense conditions provided by the external feedback that 
the first episodes of star formation in W4 provided for the eastern side of W3. 
Similarly, they also lack major sources of local internal feedback comparable 
to that of the HDL. 
Indeed, compared to the HDL, the localized star
formation and the much lower level of local feedback in the western fields
could also explain the low $T_{\mathrm{env}}$ of KR 140 and W3~NW as a
whole, as well as the lack of a second break in their mass
distributions. 

\subsection{The \textit{Herschel} View of the Star Formation History and Evolution of the W3 GMC}\label{ssec:history}

Arising from this work is compelling observational evidence of the 
dramatic effects that stellar feedback can have on the evolution of a cloud.
Initially, dense structures with conditions particularly favorable for new star formation can be created by 
local (in-cloud) or, in the case of the HDL, external feedback effects.  
These initial (past) conditions determine the properties of the first and subsequent 
episodes of star formation and associated local feedback events. 
The cumulative effects of this sequence of events will ultimately determine 
the general state of the region at the current epoch being observed.

Quantifying the balance between destructive/constructive effects of stellar feedback, 
influencing (opposing or aiding) known physical processes associated with star formation, 
such as self-gravity, accretion, and turbulence, is therefore fundamental for understanding 
the star formation process itself, from cloud to galactic scales. 
Our \herschel-based results can be used as observational constraints to construct a 
coherent evolutionary model of the W3 GMC, and therefore of 
the conditions and processes that ultimately originate the exotic high-mass 
stars and clusters.

Past studies indicate that star formation in the western fields must
have been initiated at the same time as (if not before) the HDL
(e.g., \citealp{rivera2011}).  Results from the present work suggest,
in addition, that the main difference between the evolution and
properties of the structures and stellar population of the different fields 
might be the \textit{magnitude} of the local stellar feedback occurring within 
those fields. This difference in feedback level can, however, 
itself be linked to the primordial
amount of mass and mean density in these fields, and therefore the
initial level of star-forming activity. 

A lower initial star formation efficiency (SFE) 
in a relatively low density medium would result in lower
efficiencies for inducing star formation in secondary events, and a
lower likelihood in compressing an already low column density medium
to create structures dense enough for high-mass star formation.  While
this local compression might still result in few, relatively isolated,
high-mass stars (e.g., KR 140) and some localized star formation
\citep{rivera2011}, the limited secondary star formation and low
environmental column densities ultimately restricted the star
formation capability of the western fields to a relatively low level.

The situation in  KR 140 and W3~NW differs from the 
self-enhanced process in regions with initially
enhanced column densities, like the HDL. Such a set-up would promote
an initially high SFE, resulting in more intense (constructive)
compression and triggering events. When acting on an already dense
environment, this process would lead to the formation of new high-mass stars, 
as well as a richer stellar population (e.g., \citealp{dale2007}).  
In turn, this population would then be more effective
in further compressing and confining nearby material and therefore
further enhancing the star formation activity.  

Our conclusions would agree with the results of
\citet{carpenter2000}, who found that embedded clusters in the W3
region are preferentially located in triggered regions.
The evidence presented in this work also resembles 
the ``fireworks hypothesis" presented by \citet{koenig2012}. 
In that model, the stellar feedback by high-mass stars triggers 
the formation of richer stellar populations, a self-propagating mechanism that 
can spread through the formation of new generations of high-mass stars when 
dense material is available. 
In W3, we suggest that this process 
occurs not only due to the original prime conditions created by W4 
(i.e., the HDL), but also due to the generations of high-mass stars themselves, 
which continue creating the conditions needed for the formation of 
new populations of low-mass and, especially, high-mass stars and massive clusters.

The mechanism that we invoke for high-mass star formation in dense 
environments, and more particularly, for W3 Main, was first introduced in Paper I. 
In our proposed scenario, stellar feedback creates new dense material, 
but the properties of the high-mass star population and subsequent triggering events, 
taking place within an already triggered and therefore density-enhanced region, ensures that 
the new dense structures are also (high-mass) star-forming ones. 
This is achieved in part due to the active process of mass assembly and confinement of the 
triggering mechanism, that allows for the continuation of the star-forming process. 
We note that this differs dramatically from the scenarios presented in theoretical 
models (e.g., \citealp{dale2015}). 
In these models feedback by O stars creates 
dense material, but this material is incapable of forming stars efficiently because 
it is expelled from the potential wells that facilitate collapse. Similarly, stellar feedback 
has been predicted in some cases to lower the mass of the 
new stellar population by dispersing local material and disturbing accretion,  
(e.g., \citealp{federrath2014}). 
While more detailed simulations are required to test the specific 
scenario described here at sub-parsec scales, our observations suggest that 
the triggering conditions and the dense environments can enhance the availability of 
material and aid the accretion process. The combined effects would 
ultimately lead to the unique population of high-mass stars and clusters in W3 Main.

Star formation in W3 Main
started and progressed independently from other regions (\citealp{feigelson2008};
\citealp{rivera2011}). In the former, star formation was subsequently enhanced
by the on-going low-mass activity and the local (but large scale - several parsec) triggering
effect from IC 1795 (Figure~\ref{fig:fields}; age $\sim3-5$\,Myr;
\citealp{oey2005}; \citealp{rocca2011}), itself created by the 
original superbubble activity in W4 (e.g., \citealp{oey2005}).
This process led to the first
generation of high-mass stars in the shell around this cluster, which
ultimately led to the onset of the convergent constructive feedback in
W3 Main (Paper I).


\section{$A_{\mathrm{V}}$(HB), the Second Break in the CMD/PDF: Tracing the Origin of the Birthplaces of Clusters of High-Mass Stars}\label{sec:tracing}

A second break in the mass distributions is a property unique to the
HDL fields. In this case, that of the W3 Main/(OH) field
($A_{\mathrm{V}}$(HB)$\sim95$\,mag) is observed to be at an
$A_{\mathrm{V}}$ twice that of AFGL 333.

Column densities above the $A_{\mathrm{V}}$(HB) breaks are 
associated exclusively with the two high-mass star-forming regions W3 (OH) and W3
Main, and the AFGL 333 Ridge, the only one of the three without
confirmed high-mass star formation.  While a lack of high-mass stars
in AFGL 333 might be due to a younger age (e.g., \citealp{sakai2007};
\citealp{polychroni2012}), the fact that the mass in the AFGL 333
Ridge above its $A_{\mathrm{V}}$(HB) covers an area equivalent to the
area above the same extinction in the three clumps in W3 Main/(OH)
combined, suggests that the lower number of high-mass stars and their
farther distance from the forming structure might have led to lower
densities and a smaller degree of compression and confinement.

We observe that all the structures traced by the $A_{\mathrm{V}}$(HB) contour 
coincide with those we identified in Paper I as most likely associated
with a dynamic input of material. When acting on a region with already
enhanced column densities (like the HDL or the shell around IC 1795),
boundary high-mass stars can be particularly efficient with sub-parsec
triggering (e.g., AFGL 333 and W3 Main). If the second break in the
PDF traces those structures associated with such dynamical processes
(feedback-dominated in the case of the W3 GMC, in addition to any
additional gravitational inflow of material from the local
neighborhood this amount of mass might ultimately induce), then this
feature in the PDF could act as a signpost for locating
structures with enough mass at high extinction for possible high-mass
star formation. Note, however, that having the potential for forming
high-mass stars might not necessarily translate into actual high-mass star
formation itself, as this might occur only under very specific
conditions (as discussed in Paper I). Similarly, the actual $A_{\mathrm{V}}$ 
at which the second break occurs might vary from region to region, depending 
on factors such as local environment, strength, and direction of the triggering 
event (e.g., AFGL 333 and W3 Main).

Our results link the presence of a high-extinction break in the PDF
with the effects of an external (dynamic) effect.
Based on our observations, stellar feedback appears to be the major 
dynamic process acting within and on W3, which is the reason why feedback 
has been specifically mentioned and referred to in our discussion 
as the major driver in the evolution of W3. 
This stellar feedback-based constructive process could also be applicable to 
other regions (e.g., \citealp{xu2013}).
In more general terms, conclusions from this work in terms of high-mass 
star formation requirements would hold when stellar 
feedback is replaced or aided by other external events capable of 
recreating similar conditions.
This would be in agreement with the conclusion from the HOBYS
study of Vela C by \citet{hill2011}, who suggested that the flat part
of their PDFs could be the result of constructive large scale
flows. Like the convergent constructive feedback mechanism introduced in
Paper I, convergence of flows could indeed also satisfy the
requirement of an active input of material that in Paper I we
suggested could be the key to high-mass star and cluster formation. 
A study of the applicability of this scenario to other regions within the Galaxy, and its 
associated observational evidence (e.g., PDFs), is 
currently the focus of ongoing work (Rivera-Ingraham et al. 2015; in prep.).


\section{Conclusions}\label{sec:conclusion}

The W3 GMC offers a unique opportunity to investigate the formation
process in a variety of environments. In this second study of W3 with
\textit{Herschel} HOBYS data, we have aimed to create a coherent
picture of the evolution of this GMC, analyze its large-scale
properties and structure, and further constrain the high-mass star
formation process as first described in Paper I.  This study has been
carried out by means of a comparative analysis of the fields in W3
based on the properties derived from the \textit{Herschel} column
density and temperature maps.

The W3 Main/(OH) and AFGL 333 fields show a second break in their mass
distributions. This break appears to be related to the presence of
external dynamic processes acting on the observed structures. Since
this influence is the suggested major mechanism for forming the most
massive clusters of high-mass stars (Paper I), this break could act as
an effective signpost for identifying regions suitable for possible
high-mass star formation. The actual location of this break will
depend on the local environmental conditions.

While the first break of the PDF is expected to be influenced by
various factors (e.g., gravity), we have presented evidence that
dynamic processes such as external feedback can also be responsible
for altering the location of this break. If such processes are major
players in both shifting the break in the PDF and high-mass star
formation itself (Paper I), then it could explain why high-mass
star-forming regions have a tendency to have a higher break than
low-mass star-forming regions, as observed in previous studies.

The combined evidence provided by the YSO population and the
\textit{Herschel} datasets suggest that differences in the primordial
local conditions are key for determining the evolution and
current structural and stellar properties of each field.  A high
initial surface density, mass, and column density could allow for a
higher initial SFE. The combination of a high SFE acting in an already
high density region (like the HDL), combined with the properties of
triggering as a star formation process, could result in a
self-enhancing process in which subsequent triggering events 
lead to an increase of the very structures suitable for further star
formation. This picture is supported by the anomalously high proportion of
star-forming material in W3 as traced by \textit{Herschel}, compared
to other low-mass star-forming clouds. The same events could then lead
to a richer population, a fireworks hypothesis as suggested by
\cite{koenig2012}, as well as more massive stars. The particularly
enhanced local large scale feedback observed for the W3 Main/(OH)
field could therefore explain why this is the only field with significant
high-mass star formation.

The western fields on the other hand show only moderate stellar
feedback. This state would be commonly associated with the quiet
evolution of a cloud lacking the atypical conditions provided by W4
and the HDL, which have greatly enhanced the star-forming potential of
the eastern regions of W3.

The combined effectiveness of feedback and similar dynamic processes
(e.g., constructive convergence of flows) in 1) the creation of column
density structures suitable for star formation (as shown in this
work), and 2) star formation itself (as suggested in previous
studies), could then support the scenario where star formation
progresses simultaneously with the formation of their parent
structures. This process would translate into an increasing energy
output (luminosity) of a star-forming structure (e.g., core/clump) as
the structure itself is assembled (equivalent to a diagonal evolution
in the L/M diagram). Such a model matches the scenario introduced with
the convergent constructive feedback process in Paper I.

\acknowledgements 
AR-I acknowledges support from an Ontario Graduate Scholarship and a
Connaught Fellowship at the University of Toronto. 
AR-I is currently a Research Fellow at ESA/ESAC and also acknowledges 
support from the ESA Internal Fellowship Programme.
The authors also thank the anonymous referee for very 
useful comments that have significantly improved the clarity and content of 
the paper.
This research was supported in part by the Natural Sciences and
Engineering Research Council of Canada and the Canadian Space Agency
(CSA).
SPIRE has been developed by a consortium of institutes led by Cardiff
Univ.\ (UK) and including: Univ.\ Lethbridge (Canada); NAOC (China);
CEA, LAM (France); IFSI, Univ.\ Padua (Italy); IAC (Spain); Stockholm
Observatory (Sweden); Imperial College London, RAL, UCL-MSSL, UKATC,
Univ.\ Sussex (UK); and Caltech, JPL, NHSC, Univ.\ Colorado (USA). This
development has been supported by national funding agencies: CSA
(Canada); NAOC (China); CEA, CNES, CNRS (France); ASI (Italy); MCINN
(Spain); SNSB (Sweden); STFC, UKSA (UK); and NASA (USA).
PACS has been developed by a consortium of institutes led by MPE
(Germany) and including UVIE (Austria); KU Leuven, CSL, IMEC
(Belgium); CEA, LAM (France); MPIA (Germany); INAF-IFSI/OAA/OAP/OAT,
LENS, SISSA (Italy); IAC (Spain). This development has been supported
by the funding agencies BMVIT (Austria), ESA-PRODEX (Belgium),
CEA/CNES (France), DLR (Germany), ASI/INAF (Italy), and CICYT/MCYT
(Spain).
DP is funded through the Operational Program ``Education and
Lifelong Learning'' and is co-financed by the European Union (European
Social Fund) and Greek national funds.
This research has made use of the SIMBAD database, operated at CDS,
Strasbourg, France.


\appendix


\section{Effects of (non-GMC) Foreground/Background Material on Column Density and PDFs}\label{pdf_uncorrected}

\begin{figure}[ht]
\centering
\includegraphics[scale=0.49,angle=0]{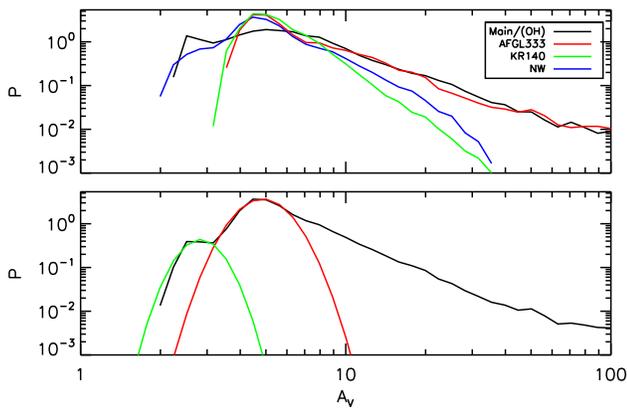}
\caption{
Top: Probability Density Functions (PDF) for each field in W3 without
correcting for the foreground/background emission. Bottom: Global PDF
for the entire W3 GMC with two fitted lognormal functions (green and
red), and a power-law tail (blue). The uncorrected PDF is
characterized by a double-peak profile.
}
\label{fig:pdf_combined_hobys}
\end{figure}

While the use of the column density maps corrected for
foreground/background material can be properly described with a single
lognormal distribution, it was observed that the use of
ISM-uncorrected maps resulted in a secondary peak at low extinctions
($A_{\mathrm{V}}\approx2.5$\,mag) for the two northernmost fields (W3
Main/(OH) and W3~NW, and therefore also on the global PDF of the
GMC). This effect can be observed in the uncorrected PDFs shown in
Figure \ref{fig:pdf_combined_hobys}.

Visual inspection of the column density maps revealed that the
material traced by this secondary peak was associated with diffuse
material \textit{external} to the W3 GMC, i.e., below average internal
environmental conditions in the GMC.  Indeed, as shown in the text,
removal of foreground/background material effectively eliminated most
of the component associated with this first peak. Only a ``remnant" of
a peak is still observed for the W3 Main/(OH) field in Figure
\ref{fig:pdf_combined_hobys}. This leftover feature is expected as our correction did
not remove contributions from the interstellar medium (ISM) local to
W3 (i.e., in the velocity range of W3 itself), adding to the shift of
pixels towards lower extinction values when removing the ISM
contribution. This example emphasizes the need for a careful selection of the
area chosen for PDF analysis. Note that this type of ``double-peaked"
PDF does not correspond to the ones observed in the vicinity of
\ion{H}{2} regions (\citealp{schneider2012}; \citealp{tremblin2014})
where the expansion of ionized gas into the molecular cloud leads to a
compressed layer of gas that shows up in the column density PDF as a
second peak at higher column densities.

While the effects on the PDF depend on the amount of ISM correction
required for a particular field, this correction is most important and
dominant at relatively low extinctions. This higher degree of
uncertainty should be taken into consideration when analyzing those
regions/structures with $A_{\mathrm{V}}<10$\,mag. Correcting for
line-of-sight material broadens and shifts the main peak of the PDFs
to lower extinctions, which change from $A_{\mathrm{V}}\sim5$\,mag to
$A_{\mathrm{V}}\sim3$\,mag. The first break of the PDFs is observed to
shift by $A_{\mathrm{V}}\sim1$\,mag, while changes to the histograms
are essentially negligible (within binning accuracy of $0.5$\,mag) for
$A_{\mathrm{V}}\gtrsim 30$\,mag.

The amount of correction needed will ultimately vary from cloud to
cloud, and even region to region. The KR 140 field, for instance,
suffers from the greatest uncertainties due to it being severely
affected by considerable foreground material traced by CO, therefore
requiring the largest correction of all fields in W3. Indeed, this
field has the largest difference in total mass before and after ISM
correction (a factor of 1.6 more mass in the uncorrected maps), and
shows the largest discrepancies in terms of mass above 
\abreak\ ($\sim3.5$ less mass in the corrected mass for the
same extinction level; Table \ref{table:sf_parameters_ism}). The mean
difference in mass above \abreak\ between corrected and
uncorrected maps for the other fields is $\sim20$\,\% of the total
corrected mass in each field, with the uncorrected maps always having
more material than the corrected ones for any given extinction
limit. This situation provides an upper limit to the mass uncertainties, as some
correction for line-of-sight-material is required. Moreover, when dealing
with uncorrected images, the \abreak\ for each field should
be higher than those derived from the ISM-corrected PDFs. Therefore,
the difference between the total mass above \abreak\ for
the uncorrected and corrected maps should be smaller than those quoted
here. None of these uncertainties affect the conclusions from this
study.


\section{Converting between CMDs and PDFs}\label{pdf_understand}

Similar to the interpretation of the break from the lognormal
distribution in the PDF (Section~\ref{ssec:break}), it has been
suggested that the point at which a break occurs in the CMD at low
extinction separates the turbulent environment from the gravity
dominated structures (\abreak;
\citealp{froebrich2010}). 

Linear regimes in the CMDs were first selected as those regions with 
slow varying gradient change, relative to those with significantly rapid change 
that should characterize a potential transition or break. 
These linear regimes were fitted in the CMDs with a line function and a $\chi^2$ minimization routine. 
The final slopes and intercepts in Table~\ref{table:fit_params} are those of the linear 
fits that best represent the data neighbouring the regions of maximum gradient 
change (closest to the breaks) as well as the overall shape of the 
CMD.
By fitting the separate extinction regimes
of the $\log$ of the CMDs with straight lines (e.g.,
Figures~\ref{fig:pdfs1} and \ref{fig:pdfs2}), these breaks can be
identified as the points where linear fits of adjacent regions
intersect. 
The sharper the transition between two regimes (with only a
small region of curvature joining adjacent linear-like regimes) the
better the constraint on the value of the break.

Considering the complexity of the PDF tails and the difficulty of
fitting them assuming the typical power-law function, in this work we
explored using the CMDs and the straightforward linear fitting method
described above as the primary approach for locating the breaks.
Because a CMD is just a cumulative form derived from a PDF, however,
any physical break should be recognizable when using either of the two
methods.

To check for consistency of breaks derived from the CMDs with features
observed in the PDFs, we transferred our best linear fits of the CMDs
($Y \equiv \log M$ vs.\ $A_{\mathrm{V}}$) to the form of the PDFs
(linear binning).  This transfer was done numerically.
Quite generally, the fraction of material (expressed here as a probability $p$) in a
given region with extinction between $A_{\mathrm{V},1}$ and
$A_{\mathrm{V},2}$ (average $A_{\mathrm{V}}$) is
\begin{equation}
p(A_{\mathrm{V}})=\frac{10^{Y_2}-10^{Y_1}}{\fmass\, \ntot\, A_{V}\,  \mathrm{Bin}}\,,
\end{equation}
where $Y_i$ is the plotted $\log$ of the CMD at $A_{\mathrm{V}_i}$,
$\fmass$ is the constant relating mass per pixel to extinction
(Section~\ref{sec:data}), $\ntot$ is the number of
valid pixels in the field, and ``Bin" is the linear bin size in
$A_{\mathrm{V}}$ used in the PDF.  This can then be transformed to the
logarithmized PDFs as displayed above.  For each line fitted to the CMDs
the corresponding colored curves are plotted in the left panels of Figures~\ref{fig:pdfs1} and \ref{fig:pdfs2}.

For the specific case in Table~\ref{table:fit_params} and
Figures~\ref{fig:pdfs1} and \ref{fig:pdfs2}, where $\log M =
\atslope\, A_{\mathrm{V}} + \log M(0)$ or
\begin{equation}
\ln M = \aslope\, A_{\mathrm{V}} + \ln M(0)\, ,
\label{mst}
\end{equation}
using $\flogs = \ln 10$ and $\aslope = - \flogs \atslope$,
the underlying PDF generating function is
\begin{equation}
p(A_{\mathrm{V}})  = \frac{M(0)\, \aslope}{A_{\mathrm{V}}\, \fmass \ntot} \exp(-\aslope A_{\mathrm{V}})\, . 
\label{epmst}
\end{equation}
When cast in the format needed for the logarithmized dependent and
independent variables used in Figures~\ref{fig:pdfs1} or
\ref{fig:pdfs2}, left, the curve has the form
$\log \plogt(\log A_{\mathrm{V}}) = [\,\log M(0) + \log (- \flogs^2 \atslope) - \log (\fmass \ntot)\,] + \atslope\, 10^{\log
A_{\mathrm{V}}}$.




\end{document}